\documentclass[12pt]{article}
\usepackage{a4wide,epsfig,amsmath,amssymb,cite,scalefnt,graphicx}
\def\prl{Phys.\ Rev.\ Lett.\ }

\def\be{\begin{equation}}
\def\ee{\end{equation}}
\def\besub{\begin{subequations}}
\def\eesub{\end{subequations}}

\def\Gbar{\overline G}

\def\pa{\partial}

\def\lambdabar{\bar\lambda}
\def\frak#1#2{{\textstyle{\frac{#1}{#2}}}}

\def\Ncal{{\cal N}}

\def\Fbar{\overline F}

\def\hbar{{\overline h}}
\def\sigmabar{{\overline\sigma}}

\def\thetabar{{\overline\theta}}
\def\alphadot{\dot\alpha}

\def\Qbar{\overline Q}
\def\psibar{\overline{\psi}}

\def\phibar{\overline{\phi}}
\def\ybar{\overline y}

\newcommand{\beq}{\begin{equation}}
\newcommand{\eeq}{\end{equation}}
\newcommand{\beqa}{\begin{eqnarray}}
\newcommand{\eeqa}{\end{eqnarray}}

\def\lsl{\not{\hbox{\kern-1.7pt $\ell$}}}
\def\ksl{\not{\hbox{\kern-2.1pt $k$}}}
\def\Psl{\not{\hbox{\kern-2.1pt $P$}}}

\def\be{\begin{equation}}
\def\bea{\begin{eqnarray}}
\def\eea{\end{eqnarray}}
\def\nn{\nonumber\\}

\newcommand{\eqn}[1]{Eq.~(\ref{#1})}
\newcommand{\eqns}[2]{Eqs.~(\ref{#1}),(\ref{#2})}

\begin{document}

\begin{titlepage}
\begin{flushright}
LTH818\\ 
\end{flushright}
\date{}
\vspace*{3mm}

\begin{center}
{\Huge
The non-anticommutative supersymmetric $U_1$ gauge theory}\\[12mm]
{\bf I.~Jack, D.R.T.~Jones and R. Purdy}\\

\vspace{5mm}
Dept. of Mathematical Sciences,
University of Liverpool, Liverpool L69 3BX, UK\\

\end{center}

\vspace{3mm}
\begin{abstract}
We discuss the non-anticommutative ($\Ncal=\frak12$) 
supersymmetric $U_1$ gauge theory in four
dimensions, including a superpotential. We perform the one-loop 
renormalisation of the model, 
including the complete set of terms necessary for renormalisability,
showing in detail how the eliminated and uneliminated forms of the
theory lead to equivalent results.
\end{abstract}

\vfill

\end{titlepage}

\section{\label{intro}Introduction}
Deformed quantum field theories have been subject to  
renewed attention in recent years due to their natural appearance 
in string theory. Initial investigations focussed on theories on 
non-commutative spacetime in which the commutator of the spacetime 
co-ordinates becomes
non-zero. More recently\cite{casal,casala,brink,schwarz,ferr,klemm,abbas,deboer,
oog}, non-anticommutative supersymmetric theories have been 
constructed by deforming the anticommutators of the Grassmann co-ordinates
$\theta^{\alpha}$
(while leaving the anticommutators of the $\thetabar{}^{\alphadot}$ unaltered).
Consequently, the anticommutators of the supersymmetry generators 
$\Qbar_{\alphadot}$ are
deformed while those of the $Q_{\alpha}$ are unchanged. It is straightforward 
to construct non-anticommutative versions of ordinary supersymmetric theories
by taking the superspace action and replacing ordinary products by
the Moyal $*$-product\cite{seiberg} which implements the non-anticommutativity.
Non-anticommutative versions of the Wess-Zumino model and supersymmetric gauge
theories have been formulated in four 
dimensions\cite{seiberg,araki} and their renormalisability
discussed\cite{brittoa,terash,brittob,rom,lunin}, 
with explicit computations up to two loops\cite{grisa} for the Wess-Zumino 
model and one loop for gauge theories\cite{jjwa,jjwb,penrom,grisb,jjwc}. 
Even more recently, non-anticommutative theories in two dimensions have been 
constructed\cite{inami, chand, chanda, luis,chandc}, and 
their one-loop divergences computed\cite{arakib,jp}.
In Ref.~\cite{jjp} 
we returned to a closer examination of the non-anticommutative
Wess-Zumino model (with a superpotential) in four dimensions, and 
showed that to correctly obtain  
results for the theory where the auxiliary fields have been eliminated,   
from the corresponding results for 
the uneliminated theory, it is necessary to include
in the classical action separate couplings for all the terms which may be 
generated by the renormalisation process. 

It seems natural to extend the above calculations to the gauged case, for
which we seek the simplest possible gauged extension of the Wess-Zumino
model with a (trilinear) superpotential.  
General gauged 
non-commutative theories were considered 
earlier\cite{jjwa,jjwb,penrom,grisb,jjwc}, and in particular
gauged interacting theories in Ref.~\cite{jjwc}; however there
we only considered a 
trilinear superpotential in the adjoint $SU_N$ case, and a 
mass term in the fundamental $U_N$ case.
 The simplest model with a 
trilinear superpotential is the three-field Wess-Zumino model with a 
$U_1$ gauge invariance, and it is this model we shall consider here. We shall
consider the one-loop renormalisation of this model in its entirety; the 
divergent contributions in the
absence of a superpotential can be extracted from Refs.~\cite{jjwa}, 
\cite{jjwb}, while even some of the contributions with a superpotential may be
extracted from Ref.~\cite{jjwc} by judicious adaptation of the results there
presented for the case of the fundamental $U_N$ case with mass terms; while
a number of the divergent contributions will require a fresh diagrammatic 
computation. We start by considering the uneliminated theory and then proceed
to compare with the results from the corresponding theory with the auxiliary 
fields eliminated.

\section{Action}
In this section we shall give
the action for an $\Ncal=\frak12$ supersymmetric $U_1$ gauge theory coupled to
chiral matter with a superpotential\cite{seiberg}\cite{araki}\cite{jjwc}.  
This is obtained by the reduction to components of the deformed, 
i.e. non-anticommutative, action in superspace.   
A $U_1$ gauge-invariant superpotential
requires at least three chiral fields; we shall take exactly three, with
scalar, fermion, auxiliary components denoted $\phi_i$, $\psi_i$, $F_i$,
$i=1,2,3$. The corresponding $U_1$ charges are denoted $q_i$, $i=1,2,3$.
For simplicity we shall consider a massless superpotential.
For convenience we split the action into kinetic and potential
terms, namely   
\be
S_0=S_{\rm kin} + S_{\rm pot}
\label{action}
\ee
where
\bea
 S_{\rm kin}&=&\int d^4x
\biggl[-\frak14F^{\mu\nu}F_{\mu\nu}-i\lambdabar\sigmabar^{\mu}
(D_{\mu}\lambda)+\frak12D^2\nn
&&-igC^{\mu\nu}F_{\mu\nu}\lambdabar\lambdabar+
\Fbar_i F_i -i\psibar_i\sigmabar^{\mu}(D_{\mu}\psi)_i
-(D^{\mu}\phibar)_i (D_{\mu}\phi)_i\nn
&&+\sqrt2gC^{\mu\nu}(D_{\mu}\phibar)_i\lambdabar\sigmabar_{\nu}\psi_i
+igC^{\mu\nu}\phibar_i F_{\mu\nu}F_i+\frak14|C|^2g^2F_i\phibar_i\lambdabar
\lambdabar \nn
&&+\sum_i \Bigl\{gq_i\phibar_i D\phi_i +i\sqrt2gq_i(\phibar_i \lambda\psi_i
-\psibar_i\lambdabar\phi_i)\nn
&&-\gamma_i C^{\mu\nu}g\left[ 
\sqrt2(D_{\mu}\phibar)_i\lambdabar\sigmabar_{\nu}\psi_i+\sqrt2\phibar_i
\lambdabar
\sigmabar_{\nu}(D_{\mu}\psi)_i+i\phibar_i F_{\mu\nu}F_i\right]
\Bigr\}\biggr],
\label{skin}
\eea
and
\bea
S_{\rm pot}
&=& -\int d^4 x\Bigl[
\{(F_iG_i-y\phi_1\psi_2\psi_3
-y\phi_2\psi_3\psi_1-y\phi_3\psi_1\psi_2)+\hbox{h.c.}\}\nn
&&+2ig\ybar C^{\mu\nu}F_{\mu\nu}\phibar_1\phibar_2\phibar_3
-\frak14y|C|^2F_1F_2F_3\Bigr],
\label{spot}
\eea
where
\be
G_1=y\phi_2\phi_3,
\ee
and similarly for $G_2$, $G_3$ (corresponding to a superpotential
$W(\Phi)=y\Phi_1\Phi_2\Phi_3$). The covariant derivative is defined by
\be
(D_{\mu}\phi)_i=(\pa_{\mu}+igq_iA_{\mu})\phi_i.
\ee
In Eq.~(\ref{skin}), $C^{\mu\nu}$ is related to the non-anti-commutativity
parameter $C^{\alpha\beta}$ by
\be
C^{\mu\nu}=C^{\alpha\beta}\epsilon_{\beta\gamma}
\sigma^{\mu\nu}_{\alpha}{}^{\gamma},
\ee
where
\bea
\sigma^{\mu\nu}&=&\frak14(\sigma^{\mu}\sigmabar^{\nu}-
\sigma^{\nu}\sigmabar^{\mu}),\nn
\sigmabar^{\mu\nu}&=&\frak14(\sigmabar^{\mu}\sigma^{\nu}-
\sigmabar^{\nu}\sigma^{\mu}),
\eea
and
\be
|C|^2=C^{\mu\nu}C_{\mu\nu}.
\ee
Our conventions are in accord with Ref.~\cite{seiberg}; in particular,
\be
\sigma^{\mu}\sigmabar^{\nu}=-\eta^{\mu\nu}+2\sigma^{\mu\nu}. 
\ee
The definition
of $|C|^2$ is similarly well-established although $C^2$ might be a preferable
notation for this quantity.

For gauge invariance of $S_{\rm pot}$ we require
\be
q_1+q_2+q_3=0,
\label{anom1}
\ee
while anomaly cancellation leads to 
\be
q_1 q_2 q_3=0
\label{anom2}
\ee
so that the allowed set of charges is  
in fact $(q, -q, 0)$. This means that in fact the most general trilinear
superpotential is in fact $W=y\Phi_1\Phi_2\Phi_3+y'\Phi_3^3$ (assuming $\Phi_3$ 
to be the neutral field). We choose, however, to retain
$W=y\Phi_1\Phi_2\Phi_3$ and to present formulae in a manner 
explicitly symmetric under $q_i$ permutations; for example  
for later convenience we denote
\be
Q=q_1^2+q_2^2+q_3^2.
\ee
Note also that it follows from 
\eqns{anom1}{anom2}\ that superpotential mass terms are allowed in general; 
however as remarked earlier we will restrict ourselves to the massless case.  

It is interesting to note that the constraints 
\eqns{anom1}{anom2}\ mean that if we set $q_1 = -q_2 = q$ and 
$y = \sqrt{2}g q$ then the undeformed theory 
has ${\cal N} = 2$ supersymmetry. 

It is easy to show that $S_0$ is invariant under
\bea
\delta A_{\mu}&=&-i\lambdabar\sigmabar_{\mu}\epsilon,\nn  
\delta \lambda_{\alpha}&=&i\epsilon_{\alpha}D+\left(\sigma^{\mu\nu}\epsilon
\right)_{\alpha}\left[F_{\mu\nu}
+\frak12iC_{\mu\nu}\lambdabar\lambdabar\right],\quad
\delta\lambdabar_{\alphadot}=0,\nn
\delta D&=&-\epsilon\sigma^{\mu}D_{\mu}\lambdabar,\nn
\delta\phi_i&=&\sqrt2\epsilon\psi_i,\quad\delta\phibar_i=0,\nn
\delta\psi_i^{\alpha}&=&\sqrt2\epsilon^{\alpha} F_i,\quad
\delta\psibar_{i\alphadot}=-i\sqrt2(D_{\mu}\phibar_i)
(\epsilon\sigma^{\mu})_{\alphadot},\nn
\delta F_i&=&0,\quad
\delta \Fbar_i=-i\sqrt2D_{\mu}\psibar_i\sigmabar^{\mu}\epsilon
-2igq_i\phibar_i\epsilon\lambda
+2C^{\mu\nu}gD_{\mu}(\phibar_i\epsilon\sigma_{\nu}   
\lambdabar).
\label{trans}
\eea

The set of terms multiplied by $\gamma_i$ are separately $\Ncal=\frak12$ 
invariant under the transformations of Eq.~(\ref{trans}); they are not
in fact produced by the reduction to components of the superspace action, 
but we have anticipated the need for them later when we renormalise the theory.
It will be sufficient to take $\gamma_i$ to consist purely of 
divergent contributions.
The $|C|^2F_1F_2F_3$ and 
$|C|^2F_i\phibar_i\lambdabar\lambdabar$ terms in Eqs.~(\ref{skin}), (\ref{spot})
are also each separately 
$\Ncal=\frak12$ invariant,
and therefore could be omitted from our action without spoiling 
the $\Ncal=\frak12$ invariance. However, once we do 
include the $|C|^2F_1F_2F_3$ and
$|C|^2F_i\phibar_i\lambdabar\lambdabar$ terms,
it is necessary for the renormalisation of the model to include all possible 
terms which may be generated, as was explained in the ungauged case in 
Ref.~\cite{jjp}.
It is easy to list these terms\cite{lunin}\cite{jjwc}. The
action has a ``pseudo R-symmetry'' under 
\be
\phi_i\rightarrow e^{-i\alpha}\phi_i,\quad
F_i\rightarrow e^{i\alpha}F_i,\quad \lambda\rightarrow e^{-i\alpha}\lambda,
\quad C^{\alpha\beta}\rightarrow e^{-2i\alpha}C^{\alpha\beta},\quad
y\rightarrow e^{i\alpha}y,
\label{psea}
\ee
$\Fbar_i$, $\phibar_i$, $\lambdabar$ and $\ybar$
transforming with opposite charges to $F_i$, $\phi_i$, $\lambda$ 
and $y$ respectively, and all other fields being
neutral; and also a ``pseudo chiral symmetry'' under 
\be
\phi_i\rightarrow e^{i\gamma}\phi_i,\quad
y\rightarrow e^{-3i\gamma}y,
\label{pseb}
\ee
$F_i$ and $\psi_i$ transforming in a similar fashion to $\phi_i$ and barred 
quantities transforming with opposite charges; the gauge fields being 
unaffected. The divergent terms which can arise subject to these invariances, 
for the massless $U_1$ case and suppressing the 
$1,2,3$ subscripts, consist of (in addition to those already present
in the action) 
\be
|C|^2 F^2\phibar^2, \quad
\ybar|C|^2 F\phibar^4, \quad
\ybar^2|C|^2 \phibar^6,\quad 
\ybar |C|^2\lambdabar\lambdabar\phibar^3.
\label{terms}
\ee
The combination
\be
\ybar^{-1}[F_1\psi_2 (C \psi_3)+F_2\psi_3 (C \psi_1)+F_3\psi_1 (C \psi_2)]
\label{fpsi}
\ee
(where $(C\psi)_{\alpha}=C_{\alpha\beta}\psi^{\beta}$)
is allowed by the above symmetries and $\Ncal=\frak12$ invariant, but we
shall see later that it is not in fact generated as a
divergence in the $U_1$ theory (at least at one loop)
if it is not already present in the classical Lagrangian, and so we choose to 
omit it. Terms of the generic form $\phibar^2\psi(C\psi)$ are allowed by the
above symmetries but it is impossible to construct an $\Ncal=\frak12$ invariant 
combination which includes them. We have included in (\ref{terms})
the appropriate factors of $\ybar$ for invariance
under the pseudo-chiral symmetry. These factors are not uniquely determined 
since $y\ybar$ is invariant under this symmetry; the choice we have made
is both concise and motivated by later considerations.  

We must include all the terms in (\ref{terms})
with their own coefficient in the action 
and therefore we are led to our complete action
\be
S=S_0+S_{\rm gen}
\ee
where $S_0$ is given in Eq.~(\ref{action}) and 
\bea
S_{\rm gen}
&=&\int d^4x\Bigl[\ybar^{-1}|C|^2
\{(k_1-\frak14y\ybar)F_1F_2F_3+k_2(F_1F_2\Gbar_3+F_2F_3\Gbar_1+F_3F_1\Gbar_2)\nn
&&+k_3(F_1\Gbar_2\Gbar_3+F_2\Gbar_3\Gbar_1+F_3\Gbar_1\Gbar_2)
+k_4\Gbar_1\Gbar_2\Gbar_3\}\nn
&&+|C|^2\left\{\left(K_1-\frak14g^2\right)F_i\phibar_i
+K_2\ybar\phibar_1\phibar_2\phibar_3\right\}\lambdabar\lambdabar\Bigr].
\label{sgen}
\eea
(It is natural to impose the same cyclic symmetry on $S_{\rm gen}$ as already
present in 
the superpotential).
The $F_1F_2F_3$ and 
$F_i\phibar_i\lambdabar\lambdabar$ terms 
are now effectively assigned an arbitrary coefficient since the fact that they
are separately $\Ncal=\frak12$ invariant 
(as are all the terms in $S_{\rm gen})$ means
there is no reason for their renormalisation to be accounted for purely by
replacing quantities in $S_0$ by the corresponding bare ones;  
$\Ncal=\frak12$ invariance will not preserve the values of their
coefficients derived from the deformed superfield action.

We use the standard gauge-fixing term
\be
S_{\rm{gf}}=\frac{1}{2\alpha}\int d^4x (\pa.A)^2
\ee
with its associated
ghost terms.  The gauge propagator is given by  
\be
\Delta_{\mu\nu}=-\frac{1}{p^2}\left(\eta_{\mu\nu}
+(\alpha-1)\frac{p_{\mu}p_{\nu}}{p^2}\right)
\ee
and the fermion propagator is
\be
\Delta_{\alpha\alphadot}=\frac{p_\mu\sigma^{\mu}_{\alpha\alphadot}}{p^2},
\ee
where the momentum enters at the end of the propagator with the undotted
index.

\section{Renormalisation}
In this section we discuss the renormalisation of the gauged 
non-anticommutative Wess-Zumino model at one loop. 

\begin{figure}[H]
\includegraphics{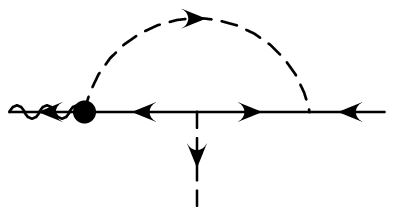}
\caption{One-loop diagram with 
a $C$ vertex and one gaugino, one $\psi$ and one $\phibar$ external legs
(a blob representing the $C$ vertex and 
dashed, full, full/wavy lines representing
scalar, fermion and gaugino fields respectively)}\label{fig8}
\end{figure}

\begin{figure}[H]
\includegraphics{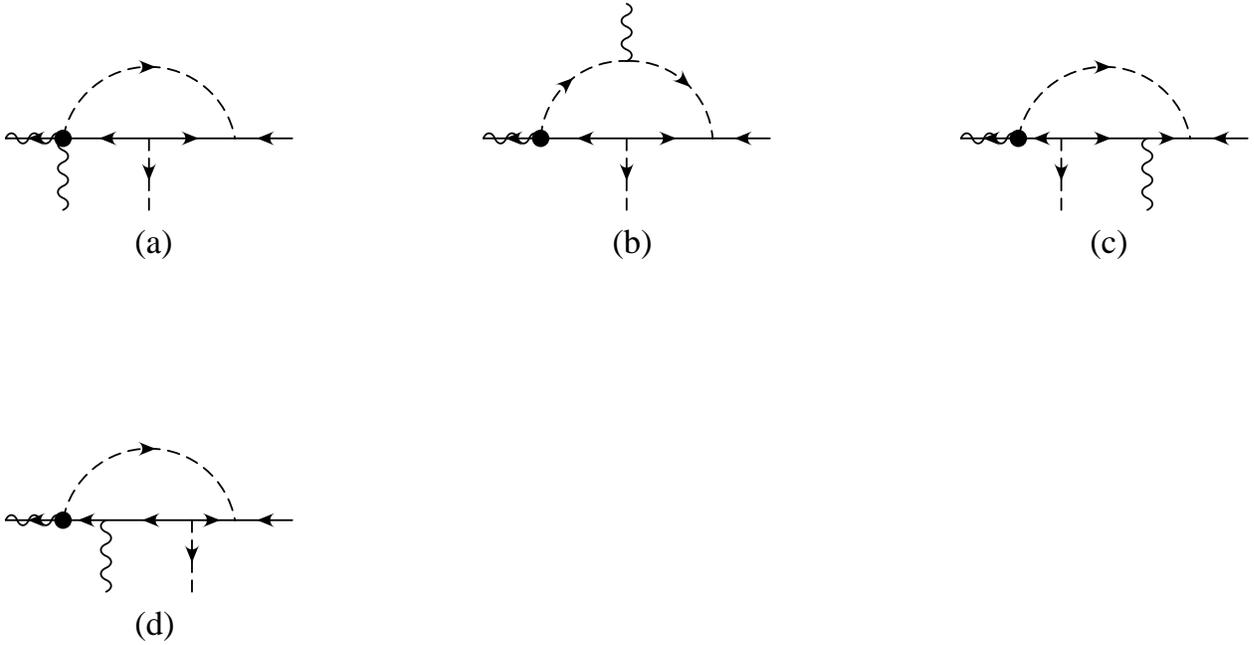}
\caption{One-loop diagrams with 
a $C$ vertex and one gauge, one gaugino, one $\psi$ and one $\phibar$ 
external legs (wavy lines reprenting gauge fields)}\label{fig9}
\end{figure}

\begin{figure}[H]
\includegraphics{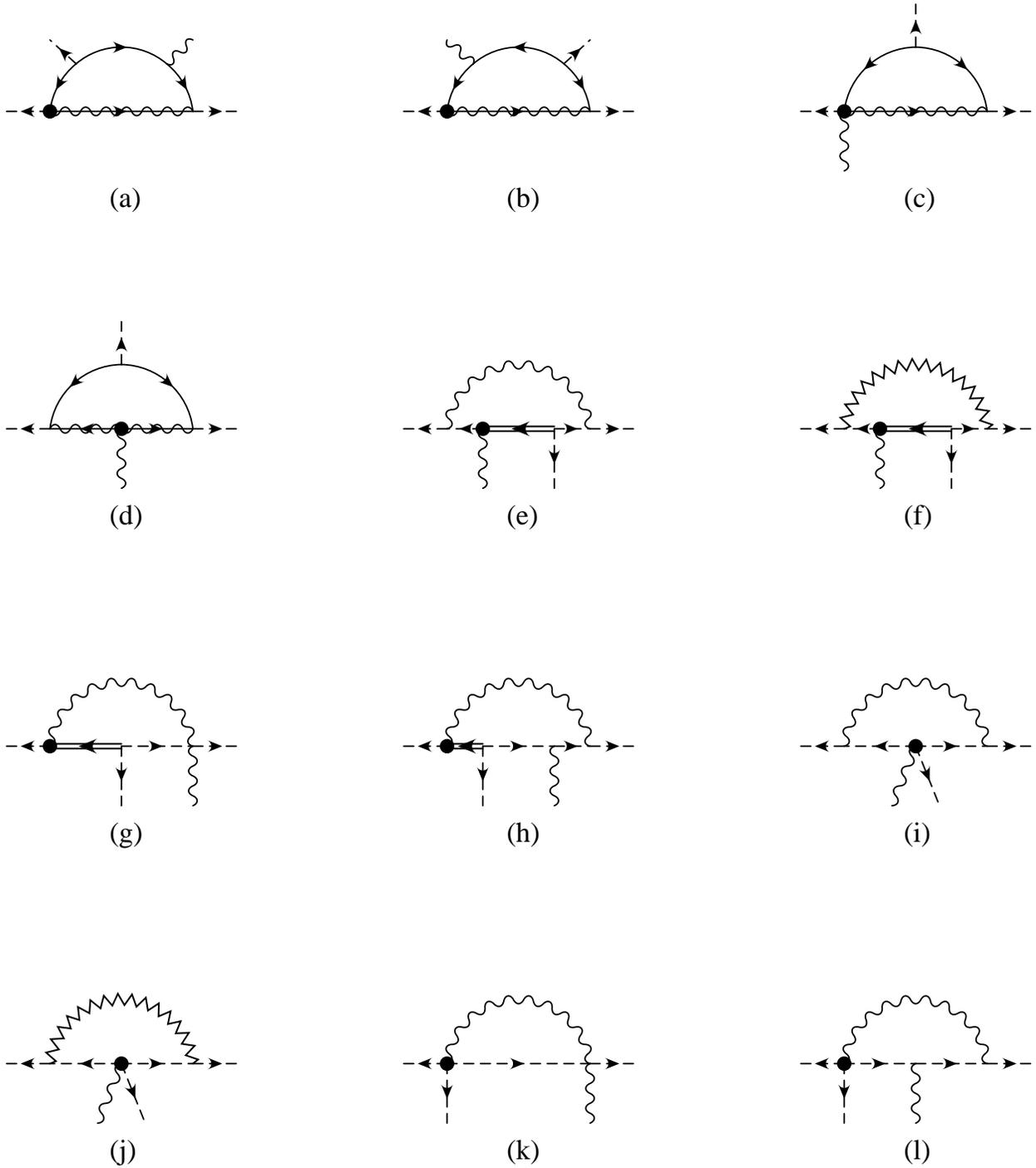}
\caption{One-loop diagrams with 
a $C$ vertex and three $\phibar$ and one gauge-field external legs
(double, zigzag lines representing
chiral and gauge auxiliary fields respectively)}\label{fig1}
\end{figure}

The divergent contributions from one-loop diagrams to terms in $S_{\rm kin}$
can mostly
be extracted from the results for the $SU_N\times U_1$ case
presented in Refs.~\cite{jjwa}, \cite{jjwb}, and so we shall just give the 
results (suppressing the well-known $C$-independent contributions)
without tabulating the contributions from individual diagrams; an exception 
is the $y\ybar$-dependent divergences, since in Ref.~\cite{jjwc}, where
we incorporated a superpotential, we did not consider the resulting
new divergent contributions to terms in $S_{\rm kin}$. The corresponding
diagrams are depicted in Figs.~\ref{fig8}, \ref{fig9}. The contribution
from Fig.~\ref{fig8} is simply
\be
-2\sqrt2y\ybar g L C^{\mu\nu}\phibar_i
\lambdabar\sigmabar_{\nu}\pa_{\mu}\psi_i,
\label{diacont}
\ee
where
\be
L=\frac{1}{16\pi^2\epsilon}.
\ee
The contributions from Fig.~\ref{fig9} are tabulated in Table~\ref{taby},
where
\be
W_1=i\sqrt2y\ybar g^2 C^{\mu\nu}A_{\mu}\sum_iq_i\phibar_i
\lambdabar\sigmabar_{\nu}\psi_i.
\ee
\begin{table}
\begin{center}
\begin{tabular}{|c| c |} \hline
a&$-2W_1$ \\ \hline
b&$W_1$\\ \hline
c&$-W_1$\\ \hline
d&$0$\\ \hline
\end{tabular}
\caption{\label{taby} Divergent contributions from Fig.~\ref{fig9}}
\end{center}
\end{table}
(In this and all the following tables the factors of $L$ are suppressed.)
Taking into account the contributions from Table~\ref{taby}, \eqn{diacont}\  
and those which can be extracted from Ref.~\cite{jjwb}, we obtain
\bea
\Gamma^{\rm pole}_{\rm kin}&=&L\int d^4x
\Bigl[-2ig^3QC^{\mu\nu}F_{\mu\nu}\lambdabar\lambdabar
-2\sqrt2gy\ybar C^{\mu\nu}\phibar_i
\lambdabar\sigmabar_{\nu}D_{\mu}\psi_i\nn
&&
+\sum_{i}\left(2\sqrt2\alpha g^3q_i^2C^{\mu\nu}D_{\mu}\phibar_i
\lambdabar\sigmabar_{\nu}\psi_i-2ig^3C^{\mu\nu}q_i^2\phibar_i F_{\mu\nu}F_i
\right)\Bigr].
\eea
The contributions to $S_{\rm pot}$, however, need to be reassessed due to the 
different form for the potential, and we therefore show the relevant 
diagrams in Fig.~\ref{fig1} and list the corresponding contributions in 
Table~\ref{tabx}.
\begin{table}
\begin{center}
\begin{tabular}{|c| c |} \hline
a&$4W_2+8W_3$ \\ \hline
b&$4W_3$\\ \hline
c&$-2W_2-12W_3$\\ \hline
d&$8W_2$\\ \hline
e&$2\alpha W_2$\\ \hline
f&$2W_2$\\ \hline
g&$-4W_2-8W_3$\\ \hline
h&$8W_3$\\ \hline
i&$-2\alpha W_2$\\ \hline
j&$-2W_2$\\ \hline
k&$4W_2+8W_3$\\ \hline
l&$-8W_3$\\ \hline
\end{tabular}
\caption{\label{tabx} Divergent contributions from Fig.~\ref{fig1}}
\end{center}
\end{table}
In Table~\ref{tabx}, $W_2$ and $W_3$ are defined by
\bea
W_2&=&iQg^3C^{\mu\nu}F_{\mu\nu}\phibar_1\phibar_2\phibar_3\nn
W_3&=&ig^3C^{\mu\nu}[q_1^2\pa_{\mu}\phibar_1\phibar_2\phibar_3
+q_2^2\pa_{\mu}\phibar_2\phibar_3\phibar_1
+q_3^2\pa_{\mu}\phibar_3\phibar_1\phibar_2]A_{\nu}.\nn
\eea
The contributions from Table~\ref{tabx} add to
\be
10iQg^3L
\int d^4 x \ybar C^{\mu\nu}F_{\mu\nu}\phibar_1\phibar_2\phibar_3.
\ee
Note that the contributions from Figs.~\ref{fig1}(e)-(h) 
cancel those from \ref{fig1}(i)-(l); we shall subsequently
omit several other pairs of diagrams where a similar
cancellation occurs (in fact we have done so already, since a potential
divergent $y\ybar C^{\mu\nu}F_{\mu\nu}F\phibar$ contribution cancels for this
reason).

\begin{figure}[H]
\includegraphics{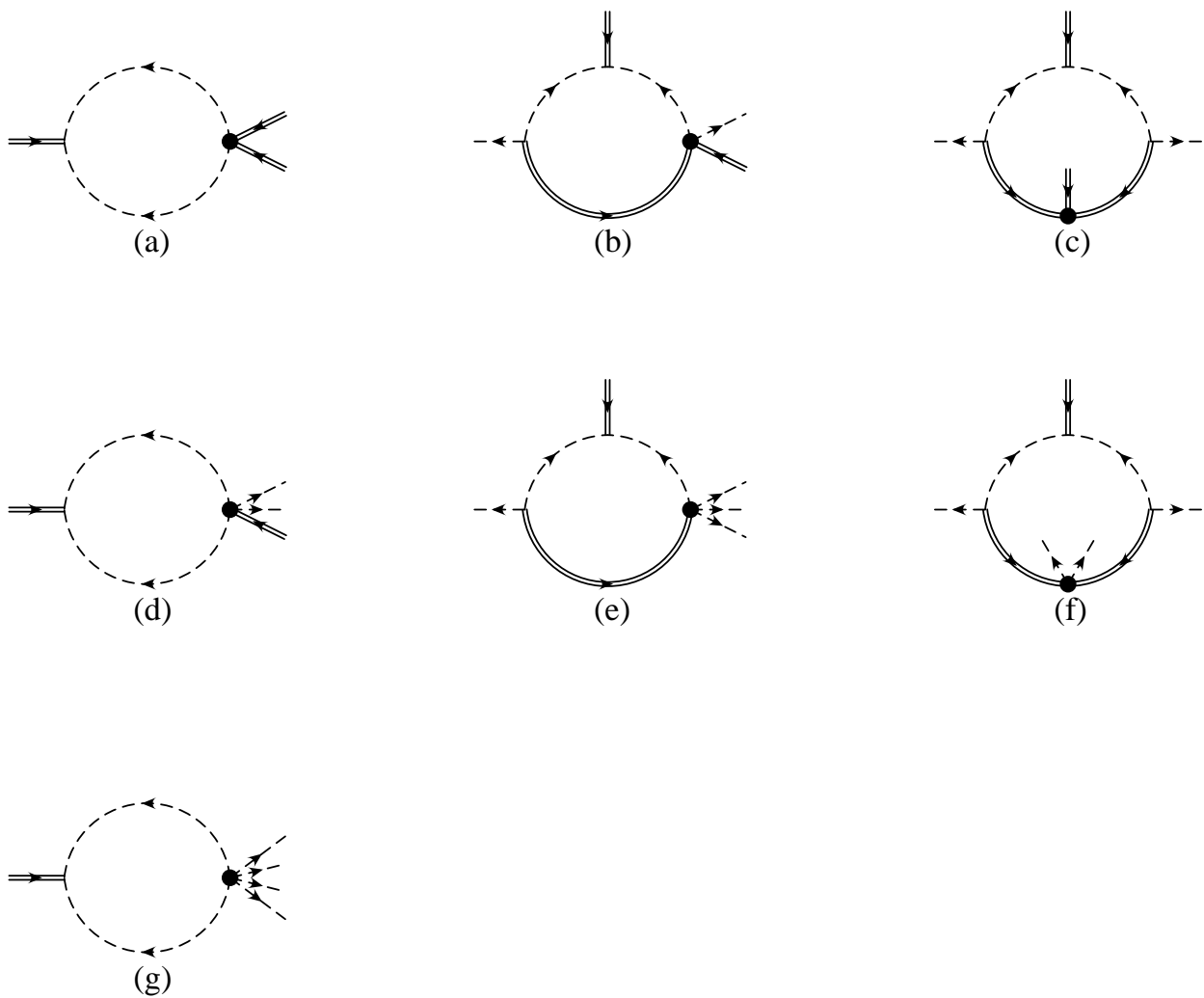}
\caption{One-loop diagrams with 
a $|C|^2$ vertex, $F$ or $\phibar$ external legs and
purely $F$ or $\phibar$ internal propagators}\label{fig2}
\end{figure}

\begin{figure}[H]
\includegraphics{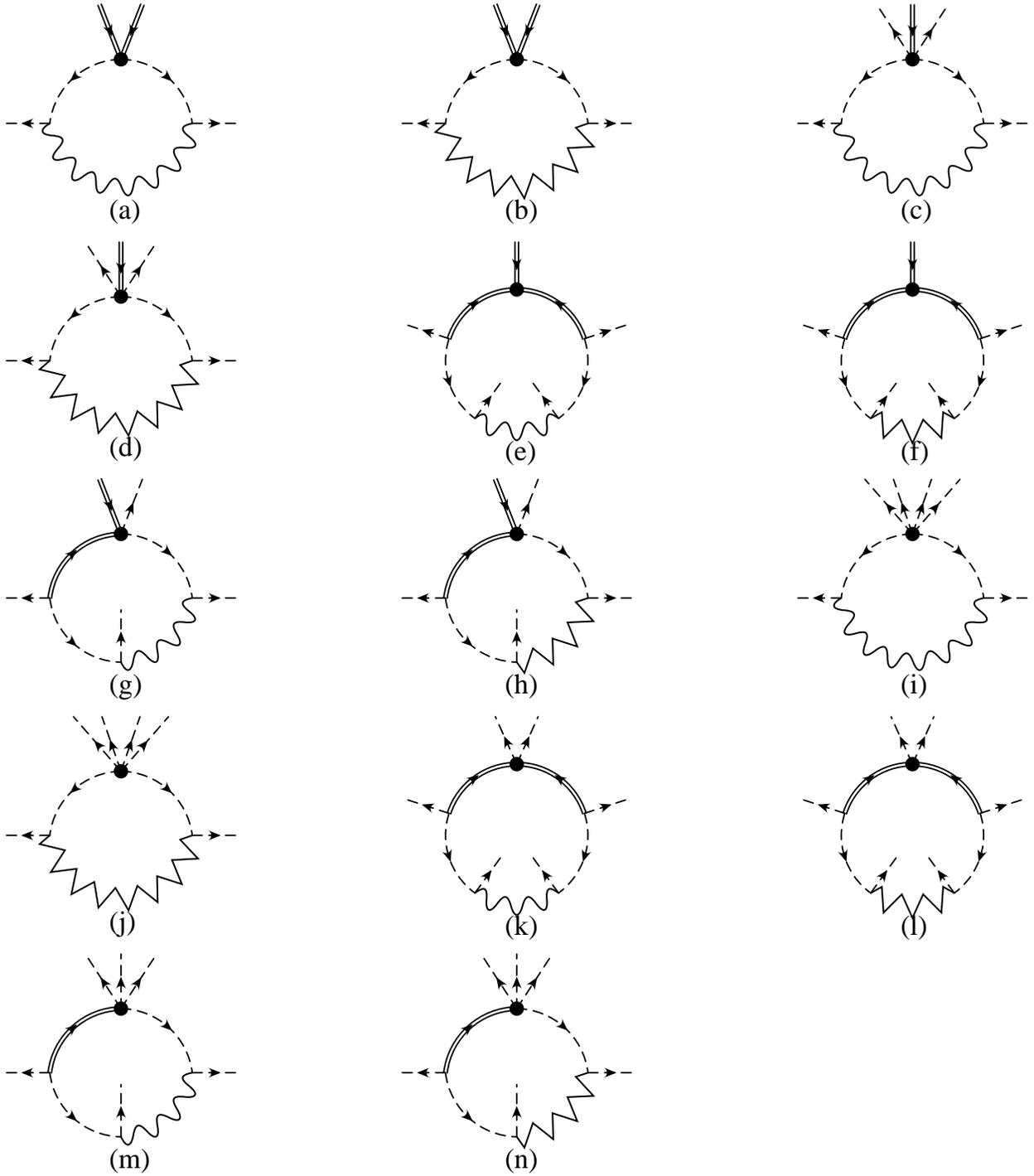}
\caption{One-loop diagrams with a $|C|^2$ vertex,
$F$ or $\phibar$ external legs   
and an internal gauge or $D$ propagator }\label{fig3}
\end{figure}

\begin{figure}[H]
\includegraphics{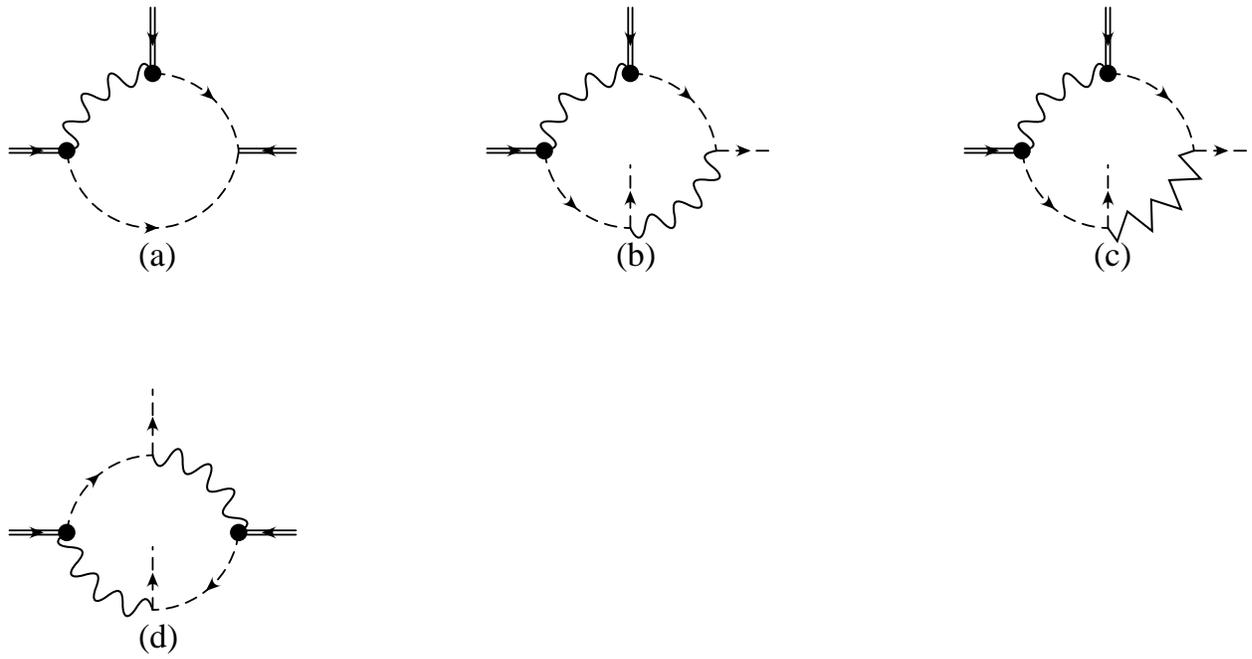}
\caption{One-loop diagrams with two $C^{\mu\nu}$ 
vertices, $F$ or $\phibar$ external legs
and an internal gauge or $D$ propagator}\label{fig4}
\end{figure}

\begin{figure}[H]
\includegraphics{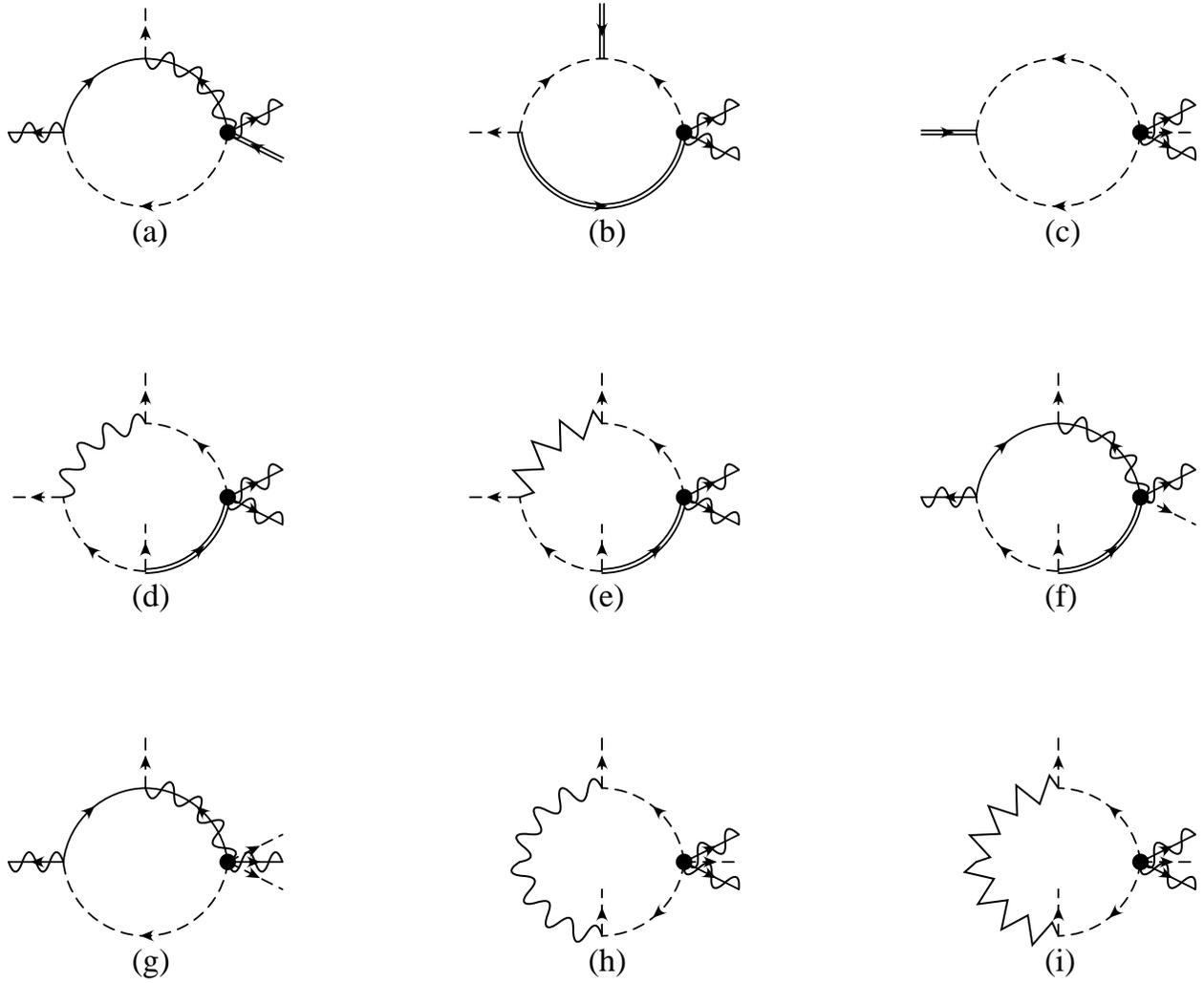}
\caption{One-loop diagrams with a $|C|^2$ vertex, 
and two gaugino and $F$ or $\phibar$
external legs }\label{fig5}
\end{figure}

\begin{figure}[H]
\includegraphics{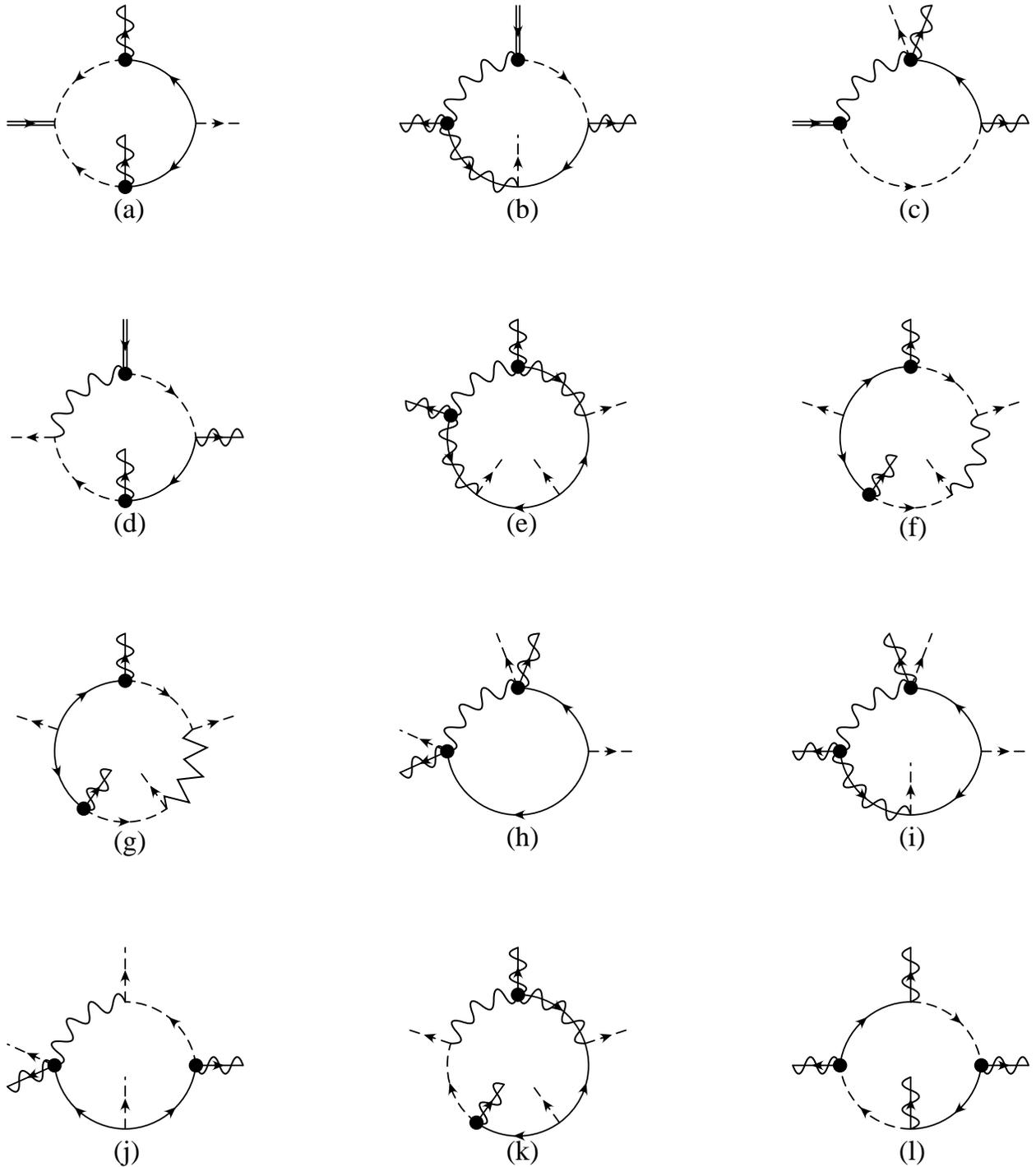}
\caption{One-loop diagrams with two $C^{\mu\nu}$ vertices,
and two gaugino and $F$ or $\phibar$
external legs }\label{fig6}
\end{figure}

\begin{figure}[H]
\includegraphics{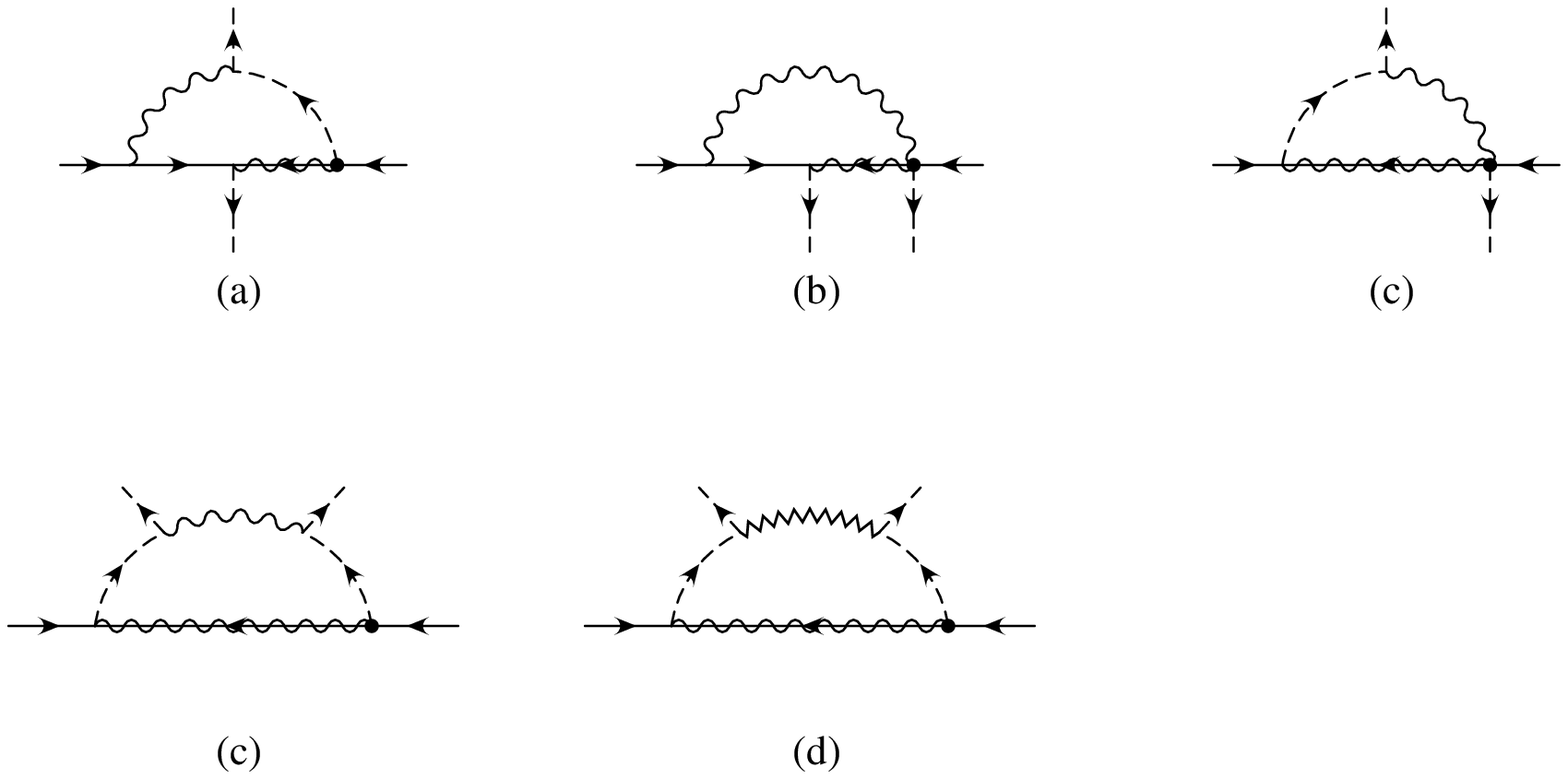}
\caption{One-loop diagrams with two $\phibar$ and two $\psi$ 
external legs (and no Yukawa vertices)}\label{fig11}  
\end{figure}

\begin{figure}[H]
\includegraphics{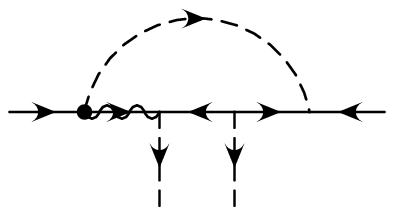}
\caption{One-loop diagram with two $\phibar$ and two $\psi$
external legs (and two Yukawa vertices) }\label{fig13}
\end{figure}

\begin{figure}[H]
\includegraphics{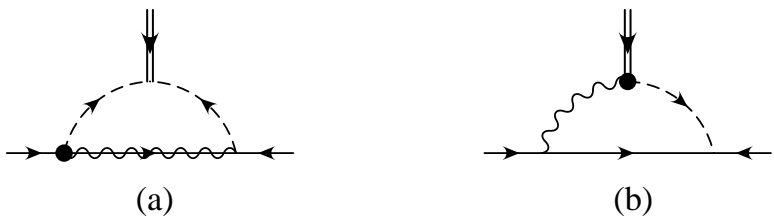}
\caption{One-loop diagrams with one $F$ and two $\psi$
external legs }\label{fig12}
\end{figure}

The divergent contributions to the $F_1F_2F_3$ and 
$F_i\phibar_i\lambdabar\lambdabar$ terms will be given in detail shortly
since these terms have now been assigned separate couplings in $S_{\rm gen}$ 
and so the divergences cannot be extracted from earlier work.  
The remaining divergent contributions are denoted by    
\bea
\Gamma^{\rm pole}_{\rm rem}&=&-\int d^4x\biggl[|C|^2\biggl\{ 
\ybar^{-1}
[X_1F_1F_2F_3+X_{2a}F_1F_2\Gbar_3+X_{2b}F_2F_3\Gbar_1+
X_{2c}F_3F_1\Gbar_2\nn
&&+X_{3a}F_1\Gbar_2\Gbar_3+X_{3b}F_2\Gbar_3\Gbar_1+X_{3c}F_3\Gbar_1\Gbar_2
+X_4\Gbar_1\Gbar_2\Gbar_3\nn
&&+X_2'(F_1^2\phibar_1^2
+F_2^2\phibar_2^2+F_3^2\phibar_3^2)+X_2''(q_1\phibar_1F_1+q_2\phibar_2F_2
+q_3\phibar_3F_3)^2]\nn
&&+\left[X_{5}F_i\phibar_i+X_5'\sum q_i^2F_i\phibar_i
+X_6\ybar\phibar_1\phibar_2\phibar_3\right]\lambdabar\lambdabar\biggr\}\nn
&&+X_7(q_1^2\phibar_1\psi_1+q_2^2\phibar_2\psi_2+q_3^2\phibar_3\psi_3)
(q_1\phibar_1C\psi_1+q_2\phibar_2C\psi_2+q_3\phibar_3C\psi_3)\biggr].
\label{divdef}
\eea
(Note the overall minus sign, introduced to avoid a proliferation of negative 
signs later on.) 
In Figs.~\ref{fig2}-\ref{fig11} are depicted the divergent one-loop diagrams 
contributing to $X_1$, etc. 
Their divergent contributions are shown diagram by
diagram in Tables \ref{taba}-\ref{tabf} and given in total by
\bea
X_1^{(1)}&=&(6k_2-6g^2)y\ybar L,\nn
X_{2a}^{(1)}&=&\{4(k_1+2k_2+2k_3)y\ybar+2(1+\alpha)k_2q_1q_2g^2\}L,\nn
X_{2b}^{(1)}&=&\{4(k_1+2k_2+2k_3)y\ybar+2(1+\alpha)k_2q_2q_3g^2\}L,\nn
X_{2c}^{(1)}&=&\{4(k_1+2k_2+2k_3)y\ybar+2(1+\alpha)k_2q_3q_1g^2\}L,\nn
X_{3a}^{(1)}&=&\{2(3k_2+6k_3+4k_4)y\ybar+
(1+\alpha)[2(k_1+2k_2)q_2q_3-Qk_3]g^2\}L,\nn
X_{3b}^{(1)}&=&\{2(3k_2+6k_3+4k_4)y\ybar+
(1+\alpha)[2(k_1+2k_2)q_3q_1-Qk_3]g^2\}L,\nn
X_{3c}^{(1)}&=&\{2(3k_2+6k_3+4k_4)y\ybar+
(1+\alpha)[2(k_1+2k_2)q_1q_2-Qk_3]g^2\}L,\nn
X_4^{(1)}&=&-(1+\alpha)(k_2+2k_3+2k_4)Qg^2L,\nn
X_2^{\prime(1)}&=&2\left(k_1+2k_2+k_3\right)y\ybar L,\nn
X_2^{\prime\prime(1)}&=&-\frak14(1+\alpha)g^4\ybar,\nn
X_5^{(1)}&=&[(4K_1+2K_2)y\ybar -g^2y\ybar)L,\nn
X_{5'}^{(1)}&=&g^2(8K_1-10g^2)L,\nn
X_6^{(1)}&=&[2(7-\alpha)K_1+(7-\alpha)K_2+14g^2]Qg^2L,\nn
X_7^{(1)}&=&16g^4L.
\label{zonea}
\eea
The terms involving $X'_2$, $X_2^{\prime\prime}$ and $X_5'$ are not 
contained in the original action; while the term involving $X_7$ is not
$\Ncal=\frak12$ invariant. However, we shall see later that all these terms may
be removed (at least at one loop) by field redefinitions.  
Other diagrams which potentially contribute divergences turn out to
be zero or to cancel. Fig.~(\ref{fig13}) is in fact zero by symmetry.
The divergences from the diagrams of Fig.~(\ref{fig12}) 
are of the form 
\be
\ybar^{-1}[(q_2-q_3)F_1\psi_2 (C \psi_3)+
(q_3-q_1)F_2\psi_3 (C \psi_1)+(q_1-q_2)F_3\psi_1 (C \psi_2)]
\ee
which (in contrast to the similar combination in (\ref{fpsi}))
is also not $\Ncal=\frak12$ invariant; moreover
there is no field redefinition which could remove these terms and so they 
must and indeed do cancel.
\begin{table}
\begin{center}
\begin{tabular}{|c| c c c c |} \hline
&$X_1$& $X_{2a,b,c}$& $X_{3a,b,c}$&$X_2'$\\ \hline
a&$6k_2y\ybar$ & &  
&\\ \hline
b&&$8k_2y\ybar$& & 
  $4k_2y\ybar$  \\ \hline
c&&$4k_1y\ybar$&&$2k_1y\ybar$\\ \hline
d&&$8k_3y\ybar$&&$2k_3y\ybar$\\ \hline
e&&&$12k_3y\ybar$&\\ \hline
f&&&$6k_2y\ybar$&\\ \hline
g&&&$8k_4y\ybar$& \\ \hline
\end{tabular}
\caption{\label{taba} Divergent contributions from Fig.~\ref{fig2}}
\end{center}
\end{table}

\begin{table}
\begin{center}
\begin{tabular}{|c| c c c c c c c|} \hline
&$X_{2a}$& $X_{2b}$&$X_{2c}$&$X_{3a}$&$X_{3b}$&$X_{3c}$&$X_4$\\ \hline
a&$2\alpha k_2q_1q_2g^2$&$2\alpha k_2q_2q_3g^2$&$2\alpha k_2q_3q_1g^2$
&&&&\\ \hline
b&$2k_2q_1q_2g^2$&$2k_2q_2q_3g^2$&$2k_2q_3q_1g^2$
&&&&\\ \hline
c&&&&$-\alpha k_3Qg^2$&$-\alpha k_3Qg^2$&$-\alpha k_3Qg^2$&\\ \hline
d&&&&$-k_3Qg^2$&$-k_3Qg^2$&$-k_3Qg^2$&\\ \hline
e&&&&$2\alpha k_1q_2q_3g^2$&$2\alpha k_1q_3q_1g^2$&$2\alpha k_1q_1q_2g^2$&
\\ \hline
f&&&&$2k_1q_2q_3g^2$&$2k_1q_3q_1g^2$&$2k_1q_1q_2g^2$&
\\ \hline
g&&&&$4\alpha k_2q_2q_3g^2$&$4\alpha k_2q_3q_1g^2$&$4\alpha k_2q_1q_2g^2$&
\\ \hline
h&&&&$4k_2q_2q_3g^2$&$4k_2q_3q_1g^2$&$4k_2q_1q_2g^2$&
\\ \hline
i&&&&&&&$-2\alpha k_4Qg^2$\\ \hline
j&&&&&&&$-2k_4Qg^2$\\ \hline
k&&&&&&&$-\alpha k_2g^2Q$\\ \hline
l&&&&&&&$-k_2Qg^2$\\ \hline
m&&&&&&&$-2\alpha k_3Qg^2$ \\ \hline
n&&&&&&&$-2k_3Qg^2$ \\ \hline
\end{tabular}
\caption{\label{tabb} Divergent contributions from Fig.~\ref{fig3}}
\end{center}
\end{table}

\begin{table}
\begin{center}
\begin{tabular}{|c| c c  |} \hline
&$X_1$&$X_2''$\\ \hline
a&$-6y\ybar g^2$& \\ \hline
b&&$-\frak14 \alpha g^4$ \\ \hline
c&&$-\frak14 g^4$ \\ \hline
d&&$0$ \\ \hline
\end{tabular}
\caption{\label{tabc} Divergent contributions from Fig.~\ref{fig4}}
\end{center}
\end{table}

\begin{table}
\begin{center}
\begin{tabular}{|c| c c c |} \hline
&$X_{5}$& $X_{6}$&$X_{5}'$\\ \hline
a&&&$8g^2 K_1$\\ \hline
b&$4K_1y\ybar$&&\\ \hline
c&$2K_2y\ybar$&&\\ \hline  
d&&$-2\alpha Qg^2K_1$& \\ \hline
e&&$-2g^2 QK_1$& \\ \hline
f&&$16Qg^2K_1$& \\ \hline
g&&$8Qg^2K_2$& \\ \hline
h&&$-\alpha Q g^2K_2$&\\ \hline
i&&$- Qg^2K_2$&\\ \hline
\end{tabular}
\caption{\label{tabd} Divergent contributions from Fig.~\ref{fig5}}
\end{center}
\end{table}

\begin{table}
\begin{center}
\begin{tabular}{|c| c c c |} \hline
&$X_{5}$& $X_{6}$&$X_{5}'$\\ \hline
a&$-g^2y\ybar$&&\\ \hline
b&&&$-8g^4$\\ \hline
c&&&$-2g^4$\\ \hline
d&&&$0$\\ \hline   
e&&$8g^4Q$&\\ \hline
f&&$\frak12\alpha Qg^4$&\\ \hline
g&&$\frak12Qg^4$&\\ \hline
h&&$\frak12(3+\alpha)Qg^4$&\\ \hline
i&&$4Qg^4$&\\ \hline
j&&$-\alpha Qg^4$&\\ \hline
k&&$0$&\\ \hline
\end{tabular}
\caption{\label{tabe} Divergent contributions from Fig.~\ref{fig6}}
\end{center}
\end{table}

\begin{table}
\begin{center}
\begin{tabular}{|c| c |} \hline
&$X_{7}$\\ \hline
a&$-4\alpha g^4$\\ \hline
b&$4(3+\alpha) g^4$\\ \hline
c&$-4\alpha g^4$\\ \hline
d&$4\alpha g^4$\\ \hline
e&$4g^4$\\ \hline
\end{tabular}
\caption{\label{tabe} Divergent contributions from Fig.~\ref{fig11}}
\end{center}
\end{table}

The divergences in Eq.~(\ref{zonea})
should be cancelled as usual by replacing the parameters
$y$, $\ybar$, $g$, $k_{1-4}$, $K_{1,2}$ 
and the fields $\phi_i$, $\phibar_i$, $F_i$, $\Fbar_i$,
$\psi_i$, $\psibar_i$, $\lambda$, $\lambdabar$
by corresponding appropriately-chosen bare 
quantities 
$y_B$, $\ybar_B$, $k_{1B-4B}$, $K_{1B,2B}$, 
$\phi_{iB}$, $\phibar_{iB}$, $F_{iB}$, 
$\Fbar_{iB}$,
$\psi_{iB}$, $\psibar_{iB}$, $\lambda_B$, $\lambdabar_B$, 
with the bare fields given by
$\phi_{iB}=Z_{\phi_i}^{\frak12}\phi_i$, etc. 
However, as emphasised in Ref.~\cite{jjwd}, renormalisation of a gauged 
supersymmetric
theory in the uneliminated case (i.e. without eliminating the auxiliary fields
$F_i$ and $D$) requires in general 
a non-linear renormalisation of $F_i$ and $D$; and in the general
$\Ncal=\frak12$ case in Ref.~\cite{jjwc} we also required a non-linear
renormalisation of the gaugino field.
In our present case we find it necessary to take at one loop
\bea
F^{(1)}_{1B}&=&Z^{\frac12(1)}_F F_1 
-(\alpha+3)q_1^2g^2\ybar L\phibar_2\phibar_3,\nn
\Fbar^{(1)}_{1B}&=&Z^{\frac12(1)}_F F_1 -(\alpha+3)q_1^2g^2yL\phi_2\phi_3
+(\alpha+9)ig^2q_1^2gLC^{\mu\nu}F_{\mu\nu}\phibar_1\nn
&&
+k_1g^2L\Bigl[\frak12(\alpha+3) (q_3^2F_2\phibar_1\phibar_2
+q_2^2F_3\phibar_1\phibar_3)
+\alpha\ybar(q_1^2-\frak12q_2^2-\frak12q_3^2)\phibar_1^2\phibar_2\phibar_3\nn
&&
+\ybar(q_1^2+\frak12q_2^2+\frak12q_3^2)\phibar_1^2\phibar_2\phibar_3\Bigr]\nn
&&+k_2g^2L\Bigl[\frak12\alpha(q_3^2F_2\phibar_1\phibar_2 
+q_2^2F_3\phibar_1\phibar_3)
+\alpha \ybar(2q_1^2-\frak12q_2^2-\frak12q_3^2)\phibar_1^2\phibar_2\phibar_3
\nn
&&-(q_1^2+q_2^2-\frak12q_3^2)F_2\phibar_1\phibar_2
-(q_1^2+q_3^2-\frak12q_2^2)F_3\phibar_1\phibar_3\nn
&&+\frak12\ybar(q_2^2+q_3^2)\phibar_1^2\phibar_2\phibar_3\Bigr]\nn
&&+k_3g^2\ybar L\Bigl[\alpha q_1^2
-(3q_1^2+2q_2^2+2q_3^2)\Bigr]\phibar_1^2\phibar_2\phibar_3\nn
&&+2(k_1+2k_2+k_3)y\ybar LF_1\phibar_1^2\nn
&&-\frak14(1+\alpha)
g^4q_1\phibar_1(q_1F_1\phibar_1+q_2F_2\phibar_2+q_3F_3\phibar_3)\nn
&&+[-10g^2+(7+\alpha)K_1]g^2Lq_i^2\phibar_i\lambdabar\lambdabar\nn
&&-\frak13Qg^2L\left[2\ybar^{-1}k_1F_2F_3+k_1(F_2\phibar_1\phibar_2
+F_3\phibar_1\phibar_3)+(2k_1-6k_3)\ybar\phibar_1^2\phibar_2\phibar_3\right]\nn
&&+\ybar^{-1}\left[R^{(1)}F_2F_3+ S^{(1)}(F_2\Gbar_3
+F_3\Gbar_2)+T^{(1)}\Gbar_2\Gbar_3\right]
\label{fren}
\eea
with similar expressions for $F^{(1)}_{2B,3B}$, and also
\be
\lambda^{(1)}_B=Z_{\lambda}^{\frac12(1)}\lambda+
i\sqrt2g\sum_i\rho^{(1)}_i\phibar_i(C\psi_i).
\label{rhodef}
\ee
Here, 
$Z_{F}$ and $Z_{\lambda}$, 
together with the renormalisation constants for the other fields 
have a loop expansion 
\be
Z_F=1+\sum_{n\ge1}Z_F^{(n)},
\ee
etc, and at one loop we have
\bea
Z^{(1)}_{\lambda}&=&-2g^2LQ,\nn
Z^{(1)}_A&=&-2g^2LQ,\nn
Z^{(1)}_g&=&g^2LQ,\nn
Z^{(1)}_{F}&=&-2Ly\ybar,\nn
Z^{(1)}_{\phi_i}&=&2L\left[-y\ybar+(1-\alpha)g^2q_i^2\right],\quad i=1,2,3\nn
Z^{(1)}_{\psi_i}&=&2L\left[-y\ybar-(1+\alpha)g^2q_i^2\right],\quad i=1,2,3.
\label{zone}
\eea
The presence of $\rho_i$ in the bare action produces terms
\be
\sum_i\rho_ig\left[\sqrt2C^{\mu\nu}( 
D_{\mu}\phibar_i\lambdabar\sigmabar_{\nu}\psi_i+\phibar_i
\lambdabar
\sigmabar_{\nu}D_{\mu}\psi_i)
+2\phibar_i\psi_i
(\sum q_j\phibar_jC\psi_j)\right]. 
\ee
The $\rho_i$ in Eq.~(\ref{rhodef}) are, like 
the $\gamma_i$ in Eq.~(\ref{skin}), purely
divergent quantities, and at one loop we find we need to take
\bea 
\gamma^{(1)}_i&=&(8g^2q_i^2-2y\ybar)L,\nn
\rho^{(1)}_i&=&8g^2q_i^2L.
\label{zoneb}
\eea
With this value for $\rho_i$, the $\Ncal=\frak12$ non-invariant terms involving
$Z_7$ in Eq.~(\ref{divdef}) are cancelled at one loop.
In Eq.~(\ref{fren}), $R$, $S$, $T$ represent possible 
additional
renormalisations of $F_i$ which are not determined by the requirements of
renormalisability.
 
With the above expression for $F^{(1)}_{iB}$, the renormalisation of the Yukawa 
couplings is as expected from applying the non-renormalisation theorem in the 
superfield context, namely
\be
y_B=\mu^{\frac12\epsilon}Z_{\Phi_1}^{-\frac12}Z_{\Phi_2}^{-\frac12}
Z_{\Phi_3}^{-\frac12}y, \quad
\ybar_B=\mu^{\frac12\epsilon}Z_{\Phi_1}^{-\frac12}Z_{\Phi_2}^{-\frac12}
Z_{\Phi_3}^{-\frac12}\ybar,
\label{nonren}
\ee 
where $\mu$ is the usual dimensional regularisation mass parameter, and 
$Z_{\Phi_i}$, $i=1,2,3$ are the renormalisation constants for the 
chiral superfields as computed in a supersymmetric gauge, namely (at one loop)
\be
Z^{(1)}_{\Phi_i}=2L\left[-y\ybar+2g^2q_i^2\right],\quad i=1,2,3.
\label{zPhi}
\ee
The $\beta$-function for $y$ is defined by 
$\beta_y=\mu\frac{d}{d\mu}y$ with a similar expression for $\beta_{\ybar}$
and then by virtue of Eqs.~(\ref{nonren}), (\ref{zPhi}),
\be
\beta^{(1)}_y=\frac{1}{16\pi^2}(3y\ybar-2g^2Q)y,
\label{betaone}
\ee
with a similar expression for $\beta^{(1)}_{\ybar}$. 

Note that  if we set $q_1 = -q_2 = q$ and 
$y = \ybar = \sqrt{2}g q$ then \eqn{betaone}\ reduces to 
\be
\beta^{(1)}_g=    2q^2\frac{g^3}{16\pi^2},
\label{betaone}
\ee
which is indeed the one-loop gauge $\beta$-function, consistent with our 
earlier remark that 
the undeformed theory has ${\cal N} = 2$ supersymmetry in this case.

We find from 
Eqs.~(\ref{sgen}), (\ref{zonea}), (\ref{fren}), (\ref{zone}), 
(\ref{zoneb}), (\ref{nonren}),  
\bea
k_{1B}^{(1)}&=&6(k_1+k_2-g^2)y\ybar L-3R^{(1)},\nn
k_{2B}^{(1)}&=&4(k_1+3k_2+2k_3)y\ybar L+R^{(1)}-S^{(1)},\nn
k_{3B}^{(1)}&=&2(k_1+5k_2+8k_3+4k_4)y\ybar L+S^{(1)}-T^{(1)},\nn
k_{4B}^{(1)}&=&3T^{(1)},\nn
K_{1B}^{(1)}&=&([6K_1+2K_2]y\ybar+2Qg^2K_1-g^2y\ybar)L,\nn
K_{2B}^{(1)}&=&2(12K_1+5K_2+2g^2)Qg^2L.\nn
\label{kone}
\eea
To a large extent the renormalisation of $\Fbar_{1,2,3}$ as given in 
Eq.~(\ref{fren}) is determined by the requirement that the couplings 
$k_{1-4}$, $K_{1,2}$ are multiplicatively renormalised as described above. 
However we still
have the freedom to choose $R^{(1)}$, $S^{(1)}$, $T^{(1)}$, which are the 
same for each 
$\Fbar_{1,2,3B}$. 
Choosing $R^{(1)}=S^{(1)}=T^{(1)}=0$ in Eq.~(\ref{fren}) leaves almost the 
minimal renormalisation of $\Fbar_i$ possible to
ensure multiplicative renormalisation; however we have 
included the terms with a factor $Q$ in Eq.~(\ref{fren})
in order to remove $g^2k_i$-dependent terms 
in $k_{1-4B}$ (something which is only possible thanks to the particular form
of the divergences, as will become clearer later when we discuss the eliminated
theory). 

Writing $\beta_{k_i}=\mu\frac{d}{d\mu}k_i$ (and similarly 
for $K_{1,2}$) and as usual requiring
that $k_{iB}$ and $K_{1,2B}$ 
be independent of $\mu$ we then find that
\bea
\beta_{k_1}^{(1)}&=&\frac{1}{16\pi^2}[6(k_1+k_2-g^2)y\ybar-3r],\nn
\beta_{k_2}^{(1)}&=&\frac{1}{16\pi^2}[4(k_1+3k_2+2k_3)y\ybar+r-s] ,\nn
\beta_{k_3}^{(1)}&=&\frac{1}{16\pi^2}[2(k_1+5k_2+8k_3+4k_4)y\ybar+s-t] ,\nn
\beta_{k_4}^{(1)}&=&\frac{3t}{16\pi^2},\nn
\beta_{K_1}^{(1)}&=&\frac{1}{16\pi^2}([6K_1+2K_2]y\ybar+2Qg^2K_1-g^2y\ybar),\nn
\beta_{K_2}^{(1)}&=&\frac{1}{16\pi^2}2(12K_1+5K_2+2g^2)Qg^2,\nn
\label{betakone}
\eea
writing $R^{(1)}=rL$, etc.
We note that these $\beta$-functions are different in form from those
derived in the ungauged case in Ref.~\cite{jjp}; of course our three-field
superpotential is also somewhat different from that used in the ungauged case,
and we have also had to include non-linear terms in $\Fbar_{1B}$ (the 
$F_1\phibar_1^2$ terms), which removed the $X_2'$ terms which would have 
spoiled renormalisability, but also contributed to $k_{3B}$. It seems
impossible to use the freedom to choose $R^{(1)}$, $S^{(1)}$, $T^{(1)}$, 
in Eq.~(\ref{fren}) to make the two sets of $\beta$-functions agree.
 
\begin{figure}[H]
\includegraphics{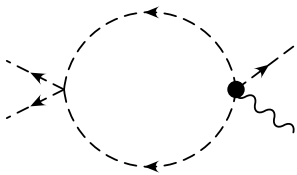}
\caption{Additional one-loop diagram for the eliminated case}\label{fig10}
\end{figure}

We now turn to the calculation in the eliminated theory.
If we eliminate $F_i$ and $\Fbar_i$ from the action we find
\bea
F_i&=&\Gbar_i,\nn
\Fbar_1&=&G_1-\ybar^{-1}|C|^2\left[k_1F_2F_3+k_2(F_2\Gbar_3+F_3\Gbar_2)
+k_3\Gbar_2\Gbar_3\right]\nn
&&-igC^{\mu\nu}F_{\mu\nu}\phibar_1
-\frak14g^2|C|^2K_1\phibar_1\lambdabar\lambdabar,
\label{felim}
\eea
(with corresponding expressions for $\Fbar_2$, $\Fbar_3$) and the action becomes
\bea
S&=&
\int d^4x\Bigl[-\frak14F^{\mu\nu}F_{\mu\nu}-i\lambdabar\sigmabar^{\mu}
(D_{\mu}\lambda)+\frak12D^2\nn
&&-igC^{\mu\nu}F_{\mu\nu}\lambdabar\lambdabar
-i\psibar_i\sigmabar^{\mu}(D_{\mu}\psi)_i
-(D^{\mu}\phibar)_i (D_{\mu}\phi)_i\nn
&&+g\sum \Bigl\{q_i\phibar_i D\phi_i +i\sqrt2gq_i(\phibar_i \lambda\psi_i
-\psibar_i\lambdabar\phi_i)\nn
&&-\gamma_i C^{\mu\nu}g\left( 
\sqrt2D_{\mu}\phibar_i\lambdabar\sigmabar_{\nu}\psi_i+\sqrt2\phibar_i
\lambdabar
\sigmabar_{\nu}D_{\mu}\psi_i\right)
\Bigr\}
+\sqrt2gC^{\mu\nu}D_{\mu}\phibar_i\lambdabar\sigmabar_{\nu}\psi_i
\nn
&&-G_i\Gbar_i+y(\phi_1\psi_2\psi_3+\phi_2\psi_3\psi_1+\phi_2\psi_3\psi_1)
+\ybar(\phibar_1\psibar_2\psibar_3
+\phibar_2\psibar_3\psibar_1+\phibar_2\psibar_3\psibar_1)\nn
&&+ig\ybar (1-\gamma_1-\gamma_2-\gamma_3)
C^{\mu\nu}F_{\mu\nu}\phibar_1\phibar_2\phibar_3
+\lambda_1\ybar^{-1}|C|^2\Gbar{}^3+\lambda_2\ybar|C|^2\phibar_1\phibar_2
\phibar_3\lambdabar\lambdabar\Bigr].
\label{selim}
\eea
where
\bea
\lambda_1&=&k_1+3(k_2+k_3)+k_4,\nn
\lambda_2&=&3K_1+K_2.
\eea
The renormalisation of the last three terms in Eq.~(\ref{selim})
now needs to be reconsidered. First
let us consider the $C^{\mu\nu}F_{\mu\nu}\phibar_1\phibar_2\phibar_3$ term. 
Its coefficient has changed, and in particular we see, 
comparing Eqs.~(\ref{spot}), (\ref{selim}), that its finite part
has changed by a factor of $-\frak12$. Moreover the diagrams
Figs.~\ref{fig1}(e)-(h) which cancelled the
contributions from Figs.~\ref{fig1}(i)-(l) are no longer present, while 
these latter contributions are multiplied by $-\frak12$. 
Moreover, since the eliminated theory 
in Eq.~(\ref{selim}) also contains
a $\Gbar_i G_i$ vertex which was not present in the uneliminated case,
there is a new diagram depicted in Fig.~\ref{fig10}, giving a divergent
contribution
\be
-6iy\ybar^2 C^{\mu\nu}\int d^4 x F_{\mu\nu}\phibar_1\phibar_2\phibar_3.
\ee
However, taking all these effects into account, it is straightforward to check
that the divergences are still cancelled. 

\begin{figure}[H]
\includegraphics{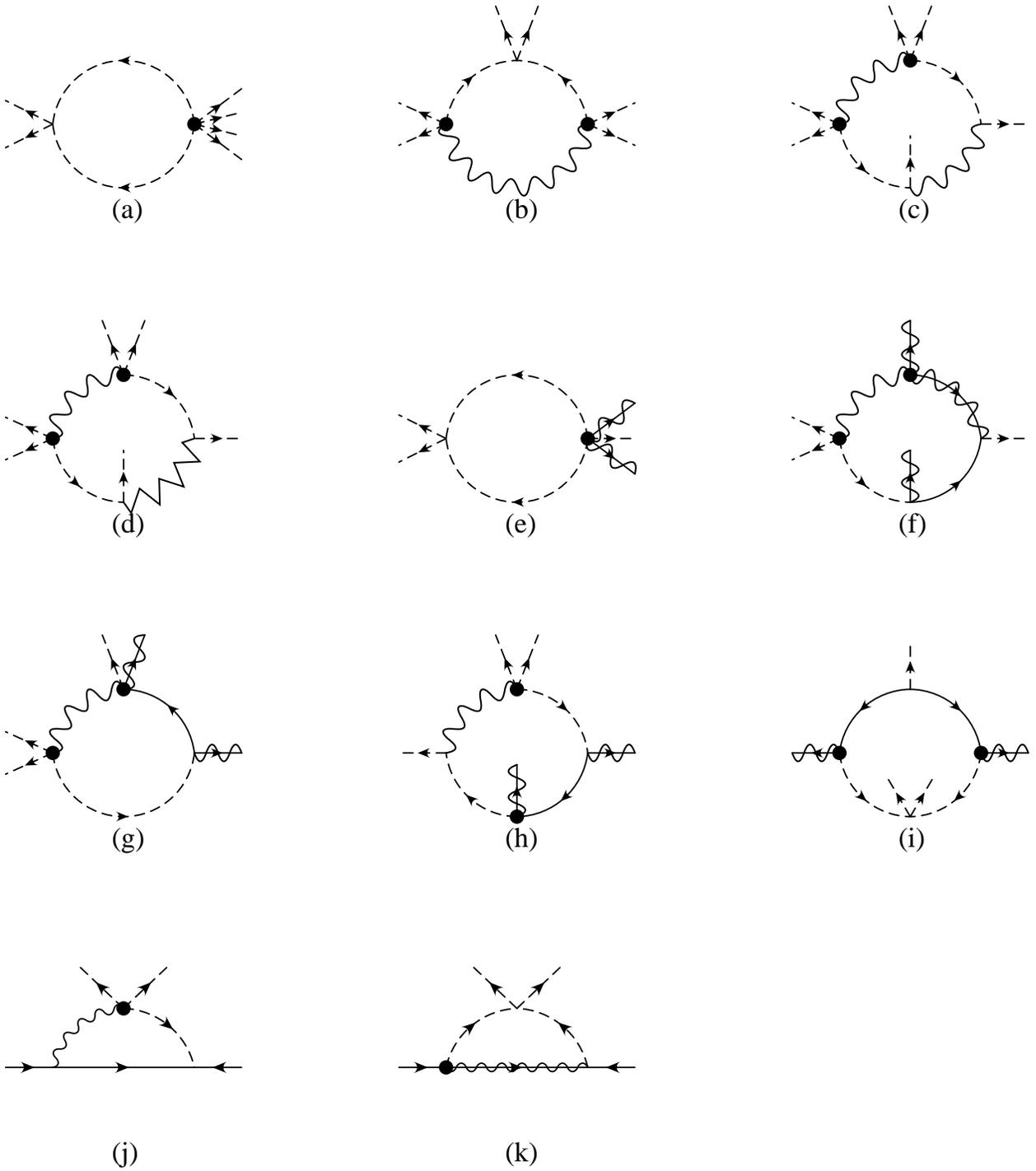}
\caption{Further one-loop diagrams for the eliminated case}\label{fig7}
\end{figure}

The remaining two terms need to be examined in more detail. 
We write the divergent contributions to these terms as
\be
\Gamma_{C\rm{elim}}^{\rm{pole}}=-|C|^2
\int d^4x[Y_1\ybar^{-1}\Gbar_1\Gbar_2\Gbar_3
+Y_2\ybar\phibar_1\phibar_2\phibar_3\lambdabar\lambdabar],
\ee
(introducing an overall minus sign as in Eq.~(\ref{divdef})).
Most of the relevant contributions to $Y_1$ can be read 
off from those to $X_4$ in Table~\ref{tabb} 
with a $k_4$ (here replaced by $\lambda_1$). Similarly,
most of the relevant contributions to $Y_2$ can be read
off from those 
to $X_6$ in Table~\ref{tabd} with a $K_2$ (here replaced by $\lambda_2$), 
and those to $X_6$ in Table.~\ref{tabe}. 
However, in the eliminated case there are also diagrams with
a $g\ybar C^{\mu\nu}F_{\mu\nu}\phibar_1\phibar_2\phibar_3$ vertex.
Such diagrams were previously cancelled by diagrams
with an internal $F$ propagator in a similar fashion to 
Figs.~\ref{fig1}(e)-(h) and \ref{fig1}(i)-(l); but of course 
such diagrams are no longer present in the eliminated case. 
Again, there are further diagrams incorporating 
the $\Gbar_i G_i$ vertex which was not present in the uneliminated case. 
The result is that we now need to 
incorporate contributions from the diagrams shown in Fig.~\ref{fig7}.
The contributions are listed in Table~\ref{tabf} (note that the contributions
from Figs.~\ref{fig7}(j), (k) cancel).

\begin{table}
\begin{center}
\begin{tabular}{|c| c c |} \hline
&$Y_1$&$Y_2$\\ \hline
a&$24y\ybar\lambda_1$&\\ \hline
b&$-6g^2y\ybar$&\\ \hline
c&$0$& \\ \hline
d&$0$& \\ \hline
e&&$6y\ybar \lambda_2$ \\ \hline
f&&$-8Qg^4$ \\ \hline
g&&$-2Qg^4$ \\ \hline
h&&$0$ \\ \hline
i&&$-3g^2y\ybar$\\ \hline
\end{tabular}
\caption{\label{tabf} Divergent contributions from Fig.~\ref{fig7}}
\end{center}
\end{table}

We find from the eliminated diagrams that
\bea
Y_1^{(1)}&=&2[12y\ybar\lambda_1-(1+\alpha)g^2Q\lambda_1-3g^2y\ybar] L,\nn
Y_2^{(1)}&=&[6y\ybar\lambda_2+(7-\alpha)Qg^2\lambda_2+4Qg^4-3g^2y\ybar] L,\nn
\label{Yelim}
\eea
and so
\bea
\beta_{\lambda_1}^{(1)}&=&\frac{1}{16\pi^2}(24\lambda_1 y\ybar-6g^2y\ybar)\nn
\beta_{\lambda_2}^{(1)}&=&\frac{1}{16\pi^2}(6y\ybar\lambda_2
+10Qg^2\lambda_2+4Qg^4-3g^2y\ybar).
\label{eq4}
\eea
An important consistency check is that
\bea
\lambda_{1B}&=&k_{1B}+k_{4B}+3(k_{2B}+k_{3B}),\nn
\lambda_{2B}&=&3K_{1B}+K_{2B},
\label{consistb}
\eea
and it is easy to confirm that this is satisfied at one loop using 
Eqs.~(\ref{kone}) and (\ref{Yelim}). The fact that we were able to remove
$g^2k_i$ terms from $k_{iB}^{(1)}$ in the uneliminated case is now seen as a 
consequence of the fact that $\lambda_{1B}^{(1)}$ contains no $g^2\lambda_1$
terms.

The original deformed Wess-Zumino action of Eq.~(\ref{action}) 
corresponded to the values $k_1=y$, $K_1=\frac14g^2$,
$k_{2-4}=K_2=0$. However,
our more general Lagrangian in 
Eq.~(\ref{sgen}) is invariant under
$\Ncal=\frak12$ transformations whatever the values of $k_{1-4}$, $K_{1,2}$;
and we see from Eq.~(\ref{betakone}) that the choice 
$k_1=y$, $K_1=\frac14g^2$, $k_{2-4}=K_2=0$
is not maintained by renormalisation;
if we set $k_1=y$, $K_1=\frac14g^2$, 
$k_{2-4}=K_2=0$ at one scale then different 
values are inevitably generated at other
scales. In Ref.~\cite{jjp} we asked (for the ungauged case) 
if there is any set of values of $k_{1-4}$ 
(or at least any form for the deformed action) 
which {\it is} preserved by renormalisation and which would be in 
some sense natural. 
Requiring that 
\be
k_i=a_i (y\ybar)^{\rho},\quad i=1\ldots4,\nn
\label{rgsol}
\ee
where $a_i$, $i=1\ldots 4$ are numbers (i.e. not functions of $y$, $\ybar$, or
$g$, and hence scale independent), entails
\be
\frac{\beta_1^{(1)}}{k_1}=\frac{\beta_2^{(1)}}{k_2}=
\frac{\beta_3^{(1)}}{k_3}=\frac{\beta_4^{(1)}}{k_4}=
\rho\left(\frac{\beta_y^{(1)}}{y}
+\frac{\beta_{\ybar}^{(1)}}{\ybar}\right).
\label{rgeq}
\ee
If we ask the same question here we shall find that the values of $k_{1-4}$
and $\rho_i$ must satisfy the sole condition
\be
[(24-6\rho)y\ybar+4\rho Qg^2]\lambda_1=6g^2y\ybar
\ee
which is the same condition we would find in the eliminated case using 
Eq.~(\ref{eq4}). In the ungauged case we once again 
find that the particular solutions
\be
k_1=-k_2=k_3=-k_4,\quad \rho=0,
\label{eq7}
\ee
and also 
\be
k_1=-\frak32k_2=3k_3,\quad k_4=0,\quad \rho=\frak13.
\label{eq8}
\ee
require no non-linear renormalisation of $F_i$.

It is tempting to feel that there is something particularly
significant about the choices in Eqs.~(\ref{eq7}), (\ref{eq8}) since they 
provided 
solutions in Ref.~\cite{jjp} at one and two loops without the need for any 
further renormalisation of $F_i$; and in fact they also solve our current model 
with the $\beta$-functions in Eq.~(\ref{betakone}), with $r=s=t=0$, i.e.
derived using the minimal renormalisation of the $F_i$ consistent with 
renormalisability.

\section{Conclusions}
We have performed a complete one-loop analysis of the 
renormalisation of the simplest gauged $U_1$ non-anticommutative Wess-Zumino 
model with a superpotential. We started with the action derived
from the non-anticommutative superspace theory, but then found 
it necessary (working with the uneliminated form of the action, 
without eliminating auxiliary fields) also to include all possible terms which 
can be generated by renormalisation with their own couplings. We showed that 
this leads to results 
compatible with those obtained in the eliminated theory. Our main results
are those in Eq.~(\ref{betakone}) (in the uneliminated case) and
Eq.~(\ref{eq4}) (in the eliminated case). This is the first complete
one-loop calculation for a general non-anticommutative supersymmetric gauge 
theory with a superpotential; as mentioned earlier, in Ref.~\cite{jjwc} we
omitted $y\ybar$ contributions to the renormalisation of terms in 
$S_{\rm kin}$.  
The renormalisation of the theory is much simpler than in the $SU_N\times U_1$
cases considered in Refs.~\cite{jjwa,jjwb,jjwc}, though once again we 
required a non-linear renormalisation of the gaugino $\lambda$, as
parametrised by $\rho_i$ in Eq.~(\ref{rhodef}), accompanied 
by a renormalisation parametrised by $\gamma_i$ in Eq.~(\ref{skin}) 
(with $\rho_i$, $\gamma_i$ as given in Eq.~(\ref{zoneb})). These 
renormalisations 
were determined by consideration of the theory with a superpotential;
however, the renormalisations contains $y$-independent pieces which
yet would not have been required in the theory without a superpotential. 
It is somewhat reassuring that the $y$-independent part of the
renormalisations for the $\rho_i$ and $\gamma_i$ is exactly as
would be obtained from the $U_1$ part of the $SU_N\times U_1$ theory of 
Ref.~\cite{jjwc}, despite the fact that here we have considered a trilinear,
three-field superpotential and there we considered a mass term 
(with two fields).

\vspace*{1em}

\noindent
{\large\bf Acknowledgements}\\
RP was supported by STFC through a graduate studentship. DRTJ was visiting
CERN and
the Aspen Center for Physics while part of this work was carried out.

\end{document}